\documentclass[amsmath,amssymb,aps,prl,reprint,superscriptaddress]{revtex4-1}
\pdfoutput=1
\usepackage{mathptmx}
\usepackage[squaren]{SIunits}
\usepackage{graphicx}
\usepackage{color}     
\usepackage{hyperref}

\newcommand{\bra}[1]{\langle #1|}
\newcommand{\ket}[1]{|#1\nolinebreak[4]\rangle}

\newcommand{\abs}[1]{\left\lvert #1 \right\rvert}

\begin{document}

\title{Spin Amplification for Magnetic Sensors Employing Crystal Defects}

\author{Marcus Schaffry}
\author{Erik M. Gauger}
\affiliation{Department of Materials, University of Oxford, Parks Road, Oxford OX1 3PH, United Kingdom}
\author{John J. L. Morton}
\affiliation{Department of Materials, University of Oxford, Parks Road, Oxford OX1 3PH, United Kingdom}
\affiliation{Clarendon Laboratory, University of Oxford, Parks Road, OX1 3PU, United Kingdom}
\author{Simon C. Benjamin}
\affiliation{Department of Materials, University of Oxford, Parks Road, Oxford OX1 3PH, United Kingdom}
\affiliation{Centre for Quantum Technologies, National University of Singapore, 3 Science Drive 2, Singapore 117543}

\date{\today}

\begin{abstract}
Recently there have been several theoretical and experimental studies of the prospects for magnetic field sensors based on crystal defects, especially nitrogen vacancy (NV) centres in diamond. Such systems could potentially be incorporated into an AFM-like apparatus in order to map the magnetic properties of a surface at the single spin level. In this Letter we propose an augmented sensor consisting of an NV centre for readout and an `amplifier' spin system that directly senses the local magnetic field. Our calculations show that this hybrid structure has the potential to detect magnetic moments with a sensitivity and spatial resolution far beyond that of a simple NV centre, and indeed this may be the physical limit for sensors of this class.
\end{abstract}

\maketitle

A key goal for future sensor technologies is the detection of weak magnetic fields at molecular length scales. There are numerous potential applications in materials science, medical science and biology. An ambitious goal  would be to detect the field due to single nuclear spins, in order to gain direct information about the structure of a molecular complex deposited on the surface. This would require unprecedented combination of sensitivity and spatial resolution.

\begin{figure}[hbtp]
  \centering
  \includegraphics[scale=0.5]{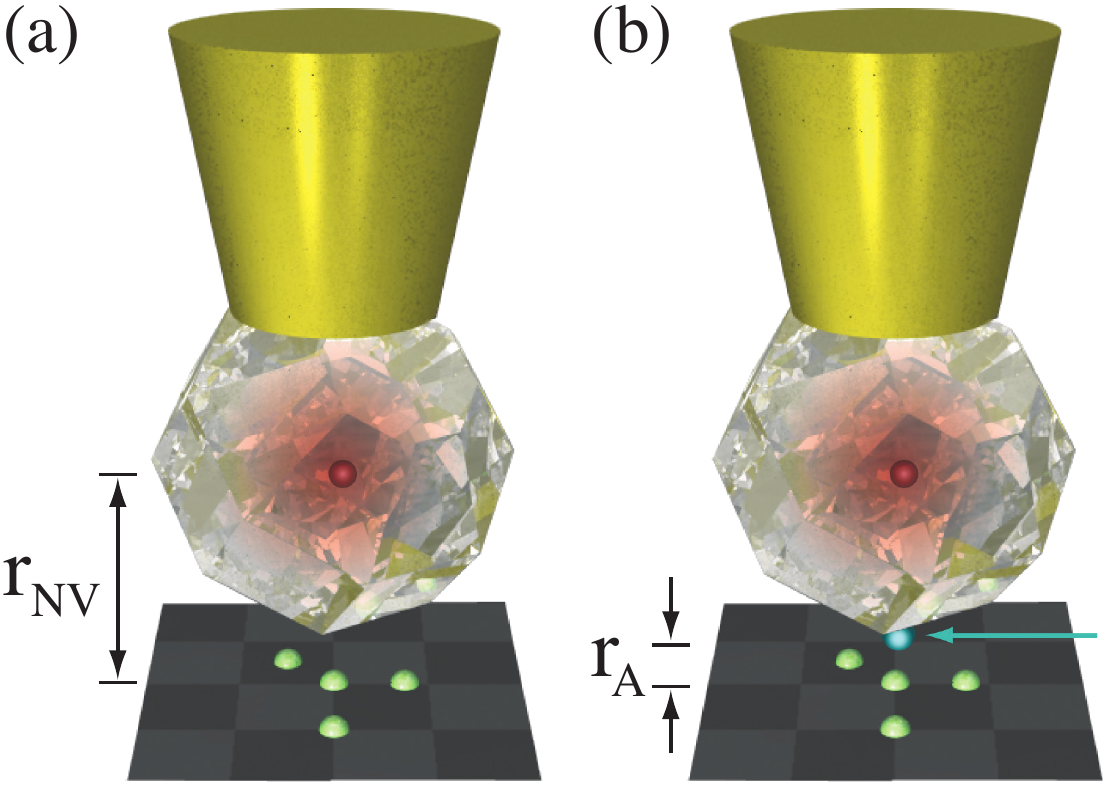}
  \caption{ (Color online) (a) The sensor structures: an NV centre (red sphere) is embedded in the middle of a diamond nano-crystal that is attached to an AFM-tip (yellow cone). The magnetic field generated by local spins (green spheres) affects the dynamic of the NV centre. The strength of the magnetic field can be inferred by measuring the NV centre through optical means and manipulating it with microwaves. (b) Amplified NV centre sensor: the surface of the nanodiamond is decorated with an another spin system (blue sphere) that couples to the NV centre inside the diamond. This additional spin system has an amplifying effect and enormously increases the magnetic field sensitivity of the sensor and its spatial resolution.}
  \label{fig:AmplifierSystem}
\end{figure}

Among the most sensitive magnetic field sensing devices are Hall sensors \cite{Ramsden2006Hallo-EffectSensors}, SQUID sensors \cite{Huber.others2008Gradiometricmicro-SQUIDsusceptometer}, force sensors \cite{Poggio.Degen2010Force-detectednuclearmagnetic}, and potentially NV centres in diamond \cite{Maze.Stanwix.ea2008Nanoscalemagneticsensing,Taylor.others2008High-sensitivitydiamondmagnetometer,Balasubramanian.Chan.ea2008Nanoscaleimagingmagnetometry,Degen2008Scanningmagneticfield,Zhao.others2011Atomicscalemagnetometryof}. In this Letter we propose a significant improvement of the NV centre sensor by coupling it to an amplifying spin system (see Fig.~\ref{fig:AmplifierSystem}). We show that the improvement in magnetic moment sensitivity can be expected to be of about three orders of magnitude; indeed the augmented NV centre may have the highest in-principle performance of any such device.

We begin with a general discussion of measuring an unknown magnetic field using two energy levels of a probe spin $s$ system ($2s \in \mathbb{N}$), the underlying principle of the NV centre sensor. Let the probe spin be governed by the Zeeman Hamiltonian $H=-\mu_{\text{Prb} } (B_0+B) S_z$, where $\mu_{\text{Prb} }$ is the magnetic moment of the probe spin, $S_z$ is the usual spin operator, $B_0$ is a known external field applied in the $z$-direction, and $B$ corresponds to the magnetic field (also in the $z$-direction) that we wish to estimate. Consider two eigenstates $\ket{0}$ and $\ket{1}$ of $H$ whose z-projections differ by an integer $m$. The field estimation then proceeds as follows: we start by creating the superposition $\ket{+}$ [here $\ket{\pm}=1/\sqrt{2}(\ket{0}\pm\ket{1})$], for example by using a suitable microwave or radio-frequency pulse sequence. This state evolves in time to $\ket{\psi(t)}=1/\sqrt{2}(\ket{0}+\exp(im\mu_{\text{Prb} }(B_0+B)t/\hbar)\ket{1})$ and therefore by successive measurements of the spin, we can infer the strength of the magnetic field. In any real-world experiment the state $\ket{\psi(t)}$ will also be affected by decoherence processes, typically dephasing is predominant  \cite{Shaji.Caves2007Qubitmetrologyand}. In this case, the coherence between  $\ket{0}$ and $\ket{1}$ decays as $\exp(-\gamma(t))$ for a positive non-decreasing function $\gamma(t)$. The time evolution of the two level system's density matrix $\rho(t)$ is then fully described by 
\begin{equation} \rho(t)=\frac{1}{2}
  \begin{pmatrix}
    1 &  e^{i\frac{m\mu_{\text{Prb} }(B_0+B)}{\hbar}t-\gamma(t)}\\
     e^{-i\frac{m\mu_{\text{Prb} }(B_0+B)}{\hbar}t-\gamma(t)}   & 1
  \end{pmatrix} \quad.
\end{equation}
The uncertainty of estimating $B$ is limited 
by the quantum Cram\'{e}r-Rao bound (CRB), ~i.e.  even for an ideal system and perfect measurements the uncertainty in $\delta B$ cannot become smaller than ${b}_{\text{CR}}$. The CRB is given by the inverse square root of the quantum Fisher information $F$ with respect to $B$ \cite{Braunstein.Caves1994Statisticaldistanceand,*Paris2009Quantumestimationquantum}, yielding the following inequality
\begin{equation}
  \delta B \geq {b}_{\text{CR}}=\frac{1}{\sqrt{F}} = \frac{e^{\gamma(\tau)}\hbar}{m\tau\abs{\mu_{\text{Prb}}}} \quad,
\end{equation}
where $\tau$ is the duration for which the spin has experienced the magnetic field. Repeating the measurement many times gives a further reduction in the uncertainty of the unknown field. Assuming the preparation of the $\ket{+}$ state takes the time $t_p$, the lower bound for the uncertainty $\delta B$ after $N=T/(\tau+t_p)$ repetitions within a total time $T$ is given by
\begin{equation}
\label{eq:CRB}
  b_{\text{CR}}=\frac{1}{\sqrt{NF}} = \frac{e^{\gamma(\tau)}\hbar\sqrt{\tau+t_p}}{\sqrt{T}m\tau\abs{\mu_{\text{Prb}}}} \quad.
\end{equation}
As an example, we set $\gamma(t)=\frac{t}{T_2}$. Minimizing the CRB with respect to $\tau$ yields the optimal sensing time for each run of the protocol:
\begin{equation}
\label{eq:linear-opt}
   \tau^*= \frac{1}{4} \left(T_2-2 t_p+\sqrt{T_2^2+4 t_p \left(3 T_2+t_p\right)}\right)   \quad.
\end{equation}
It is easily seen that $\tau^*\approx \frac{1}{2}T_2$ in the limit of weak dephasing ($T_2\gg t_p$), in contrast to $\tau^* \approx T_2$ when the dephasing time is comparatively short, $T_2 \ll t_p$.

Since a magnetic field smaller than $b_{\text{CR}}$ cannot be resolved, the CRB provides a natural limit of the sensitivity of a magnetic field sensor. Interestingly, this limit does not depend on the actual value of $B$. However, a na\"ive intuition is that the proximity of the probe spin and the sample will be an important characteristic of a practicable sensor, since it determines the strength of the measured magnetic field as well as the sensor's spatial resolution. To quantify this intuition, one can ask the question: how long does it take until the sensor spin detects the magnetic moment associated with an electron or a nuclear spin? In the following, we introduce a magnetic moment sensitivity (in units of \unit{}{\tesla \per \sqrt{\hertz}}) as a suitable figure of merit for this question: the magnetic moment sensitivity $S$ of a sensor is given by the number of proton magnetons that can be resolved within the time window $\sqrt{T}$ for a distance $r$  between the probe spin and the sample. The dipole field originating from a proton with magnetic moment $\mu_p$ at the position of the probe spin is of order $b=\frac{\mu_0}{4\pi} \frac{2\mu_p}{r^3}$ (where $\mu_0$ is the vacuum permeability), so we obtain
\begin{equation}
\label{eq:def-sen}
   S = \frac{\min_{\tau} (b_{\text{CR}})\sqrt{T}}{b} \quad.
\end{equation}
For example, once again assuming $\gamma(t)=\frac{t}{T_2}$ and employing Eq.~\eqref{eq:linear-opt} gives the following magnetic moment sensitivity
\begin{equation}
   S = \frac{2\pi \hbar}{\mu_0\mu_p}\frac{e^{\frac{\tau^*}{T_2}}\sqrt{\tau^*+t_p}r^3}{m\tau^*\abs{\mu_{\text{Prb} }}} \quad.
\end{equation}

We shall now briefly discuss the example of field sensing with an NV centre before describing the inherent limitations of this approach and presenting our proposal for improving the characteristics of this class of device. An NV centre in diamond possesses spin $1$ with the three levels $\ket{0}$ and $\ket{\pm 1}$. The levels $\ket{\pm 1}$ are degenerate in the absence of an external field and shifted by a zero-field-splitting (ZFS) of about \unit{2.87}{\giga\hertz}. For the purpose of sensing the magnetic field, various two level manifolds can be used, for example $\{\ket{0},\ket{\pm 1}\}$ or $\{1/\sqrt{2}(\ket{1}+\ket{-1}), (1/\sqrt{2}(\ket{1}-\ket{-1})\}$ \cite{Taylor.others2008High-sensitivitydiamondmagnetometer}.

\begin{figure}[hbtp]
  \centering
  \includegraphics[width=0.9\linewidth]{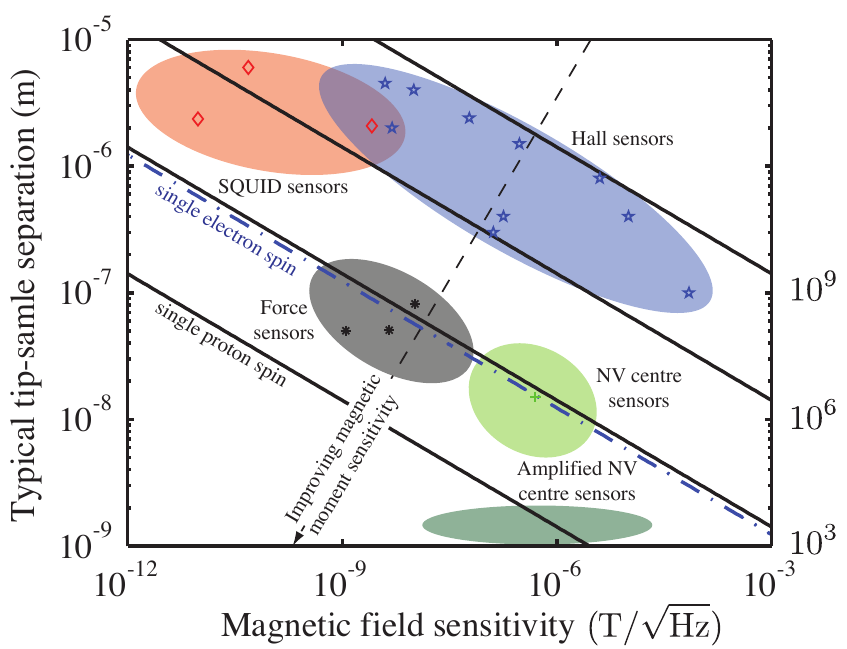}
  \caption{(Color online) Performance of various state-of-the-art and proposed magnetic field sensors (figure adapted from Ref. \cite{Degen2008NanoscalemagnetometryMicroscopy}). The plot shows magnetic sensitivity (per $\unit{}{\sqrt{\hertz}}$) (horizontal) versus a typical tip-sample separation. The data points are reported sensitivities from experiments: force sensors \cite{Rugar.Budakian.ea2004Singlespindetection,Degen.others2009Nanoscalemagneticresonance,Mamin.Oosterkamp.ea2009Isotope-SelectiveDetectionand}, Hall sensors \cite{Boero.Demierre.ea2003Micro-Halldevicesperformance}, SQUID sensors \cite{Kirtley.others1995HighresolutionscanningSQUID,Gardner.others2001Scanningsuperconductingquantum,Huber.others2008Gradiometricmicro-SQUIDsusceptometer}, and NV centre sensors \cite{Maze.Stanwix.ea2008Nanoscalemagneticsensing}. The diagonal lines sketch the boundaries for the magnetic moment sensitivity (see Eq.~\eqref{eq:def-sen}) required to detect $1,10^3,10^6,$ or $10^9$ protons (solid black) and a Bohr magneton (blue dash dotted) within one second. The potential of our proposed pre-amplified NV centre is illustrated by the dark green ellipse.}
  \label{fig:magnetic moment sensitivity}
\end{figure}

Figure~\ref{fig:magnetic moment sensitivity} shows the magnetic moment sensitivity of various classes of sensors. Clearly, NV centres hold the promise of a high field sensitivity combined with a fairly small probe to sample separation. However, in order to further improve the spatial resolution and obtain an even better magnetic moment sensitivity (according to the definition in Eq.~\eqref{eq:def-sen}) the NV centre must be brought even closer to the sample. Due to the cubic dependence of $S$ on $r$ even a modest reduction in the separation leads to a large gain, and for this reason NV centres have been embedded in smaller and smaller nano-crystals \cite{Balasubramanian.Chan.ea2008Nanoscaleimagingmagnetometry,Maze.Stanwix.ea2008Nanoscalemagneticsensing}. However, the size of the diamond crystal surrounding the NV centre cannot become arbitrarily small without severely affecting the remarkable coherence time (and thus the sensitivity) of the NV centre. In $^{12}$C isotopically enriched bulk diamond the room temperature coherence time can be as long as  \unit{1.8}{\milli\second} \cite{Balasubramanian.others2009Ultralongspincoherence}, increasing up to \unit{2.4}{\milli\second} with dynamic decoupling  \cite{Naydenov.others2011Dynamicaldecouplingof}. On the other hand, the best reported coherence time for a nanodiamond with a diameter of \unit{7}{\nano\meter} is reduced to \unit{1.4}{\micro\second} \cite{Tisler.others2009FluorescenceandSpin}. Smaller crystals with a diameter of only \unit{5}{\nano\meter} have been studied in Ref.~\cite{Bradac.others2010Observationandcontrol}, however the properties of enclosed NV centres are even further degraded in this case. A miniaturisation of the crystal hence leads to a reduced field sensitivity in exchange for a smaller separation between the probe and the sample.

We propose to overcome this trade-off situation by bringing the NV centre ''effectively'' closer to the sample without reducing the size of the nanocrystal. This is achieved by attaching an additional spin system on the surface of a nanocrystal which relays the sample magnetic field to the NV centre (see Fig.~\ref{fig:AmplifierSystem}). In the simplest implementation, a single electron spin serves as the amplifier. Of course, further improvements will be possible by using higher spin systems such as N@C$_{60}$ \cite{Morton.others2006Electronspinrelaxation} or a molecular magnet \cite{Mannini.others2010Quantumtunnellingof,Christou.Gatteschi.ea2000Single-MoleculeMagnets}. It may even be possible to make use of more sensitive entangled states with suitably engineered molecular systems \cite{Jones.Karlen.ea2009MagneticFieldSensing,Schaffry.others2010Quantummetrologywith}.

\begin{figure}[hbtp]
  \centering
  \includegraphics[width=0.9\linewidth]{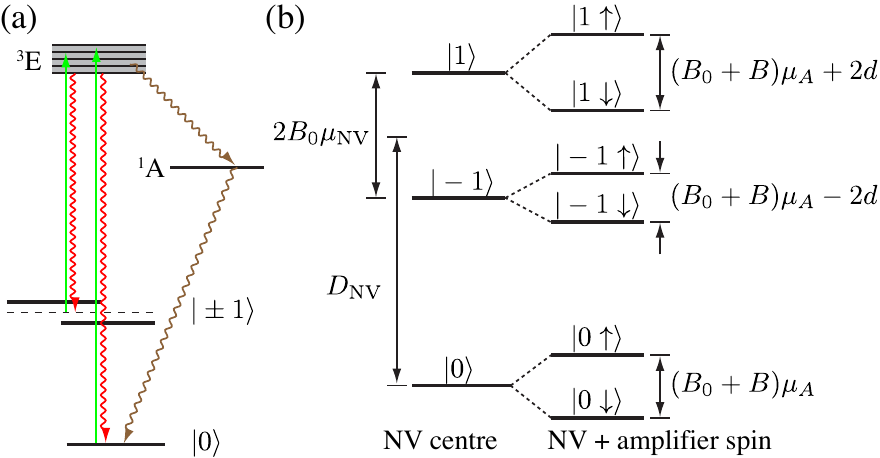}
  \caption{(Color online) (a) Level structure of an NV centre. The ground state is a spin triplet with zero field splitting and the excited state has a manifold of levels. Following excitation with green light the dominant decay is spin preserving. However, spin is not always conserved due to an additional decay path to $|0\rangle$ via a metastable state, so that continuous excitation eventually results in a polarised state. (b) Eigenspectrum of $H$ for an amplifier spin $\frac{1}{2}$ system.}
  \label{fig:eigenspectrum}
\end{figure}

For simplicity, we consider an electron spin as the amplifier in the following. The NV centre and the amplifier spin are dipolar coupled, and we assume the vector connecting the two spins is aligned with the $z$-axis. Both spins experience an (known) homogeneous external magnetic field $\mathbf{B}_0 = (0,0, B_0)$ and a small magnetic field whose $z$-component $B$ at the position of the amplifier is to be measured. In comparison to the amplifier spin the NV centre experiences a weaker sample field $c B$ where the factor $c < 1$ depends on the relative separations. The Hamiltonian thus reads:
\begin{multline}
 H = - \mu_{\text{NV}} (B_0+c B)S_{\text{NV},z} + D_{\text{NV}} S_{\text{NV},z}^2 - \mu_A (B_0+B) S_{A,z}\\ 
+  d( 2  S_{\text{NV},z} S_{A,z}-  S_{\text{NV},x} S_{A,x}-  S_{\text{NV},y} S_{A,y} ) \quad,   
\end{multline}
where $d=\frac{\mu_0}{4\pi} \frac{\gamma_{\text{NV}} \gamma_A\hbar}{\Delta^3}$ is the dipolar coupling constant; $\gamma_{\text{NV}}$ and $\gamma_A$ are the gyromagnetic ratios of the spins, $\Delta$ is the distance between the NV centre and the amplifier spin, and $D_{\text{NV}}$  denotes the NV centre's ZFS constant. Our protocol assumes that the influence of $cB$ on the NV centre is much smaller than its coupling to the amplifier spin; this will certainly be justified if the amplifier is an electron spin and the sample field is due to nuclear spins. In addition we assume that flip-flops between the spins are heavily suppressed and can be neglected, for example because $D_{\text{NV}}$ represents the largest energy in the system, yielding the following effective Hamiltonian
\begin{multline}
 H \approx - \mu_{\text{NV}} B_0 S_{\text{NV},z} + D_{\text{NV}} S_{\text{NV},z}^2  \\ 
 -\mu_A (B_0+B) S_{A,z} + 2d S_{\text{NV},z} S_{A,z} \quad,
\end{multline}
whose level structure is schematically depicted in Fig.~\ref{fig:eigenspectrum}.

The field sensing protocol now proceeds as follows. Initially, the system starts in a completely mixed spin state, $\frac{1}{6}\mathbf{1}_6$, so that it needs to be polarised with laser and microwave pulses. As a first step we polarise the NV centre by illuminating it with green light, after about \unit{1}{\micro\second} the NV centre will have relaxed to the state $\ket{0}$ with high probability \cite{Taylor.others2008High-sensitivitydiamondmagnetometer}, leaving the combined system in the state $\ket{0}\bra{0} \otimes \frac{1}{2}\mathbf{1}_2$. Next, we apply a selective microwave $\pi$-pulse on the transition $\ket{0}\ket{\downarrow}\leftrightarrow\ket{1}\ket{\downarrow}$. For a distance $\Delta=\unit{10}{\nano\meter}$ between the NV centre and the amplifier, this transition is split from the neighbouring $\ket{0}\ket{\uparrow}\leftrightarrow\ket{1}\ket{\uparrow}$ transition by $2d=2\frac{\mu _0 g_{\text{NV}}g_A\mu _B^2}{4 \pi \Delta^3\hbar}\approx \unit{0.65}{\mega\rad\per\second}$ (i.e.~$\unit{104}{\kilo\hertz}$). A highly selective $\pi$-pulse is beneficial for the presently discussed protocol. For this reason we now establish the conditions under which imperfections in the pulse selectivity can be considered negligible. Assuming Lorentzian line broadening of both transitions with a FWHM of $2/T_{2,\text{NV}}$ ($T_{2,\text{NV}}$ is the coherence time of the NV centre), a detailed analysis 
\footnote{Reduced $\pi$-pulse selectivity gives rise to two types of error: firstly, a mixed initial state, resulting in a correction factor of $(1+p)/(1-p)$ for the sensitivity $S_{A}$, where $p$ describes the proportion of the population flipped in the unwanted transition. Secondly, a degradation of the Fisher information entailing an additional correction factor of $\sqrt{(1+p)/(1-p)}$. Both factors are very close to 1 as long as the pulse remains highly selective.}
shows that a linewidth overlap of up to  $10 \%$ meets this requirement and translates into $T_{2,\text{NV}} > \unit{10}{ \micro \second}$. The product of pulse duration and frequency passband $T \times \Delta \omega$ is typically between 2 and 5 \cite{Bernstein.King.ea2004HandbookofMRI}. Taking $T = 4/ \Delta \omega$ and allowing the passband to overlap with the area of unwanted transition's Lorentzian by at most $5\%$ (allowing us to neglect imperfect pulse selectivity), we thus obtain $\tau_{\pi} = \frac{4}{2(2d-\cot(0.05\pi)/T_{2,\text{NV}})}$ for the duration of the desired selective $\pi$-pulse (reducing to $\tau_{\pi} \approx 1/d$ in the limit of long $T_{2,\text{NV}}$). Finally, the state of the NV centre is measured by exciting it once more with green light and detecting the spin-dependent photoluminescence. Importantly, this puts the NV centre back into the $\ket{0}$ state, leaving the system in one of the well-defined states $\ket{0}\ket{\downarrow}$ or $\ket{0}\ket{\uparrow}$.

Without loss of generality we assume the system is in the state $\ket{0}\ket{\downarrow}$, ready for the magnetic field estimation through several repetitions of the following sensing cycle: A microwave $\frac{\pi}{2}$-pulse creates the superposition $1/\sqrt{2}(\ket{0\downarrow}+\ket{0\uparrow})$. It is safe to assume that the amplifier spin can be rotated fast and with high fidelity as long as the unknown field corresponds to the smallest energy scale of the system, as can be accomplished with a modest external field. After a sensing time $\tau$, the amplifier spin acquires a relative phase proportional to $B$. The phase is first mapped onto a population difference between $\ket{0\downarrow}$ and $\ket{0\uparrow}$ with another $\frac{\pi}{2}$-pulse, and then entangled with the NV centre spin state with a selective $\pi$-pulse in the same way as in the initialisation process. The NV centre spin state is now read out using spin-dependent photoluminescence detection (also initialising it for the next sensing cycle).

Having described the protocol, we now benchmark the sensitivity improvement of the amplified system over a conventional single NV centre sensor. Obviously, the performance of both sensors depends on the decoherence model, which we assume to be fully characterised by $\gamma(t)$. Different forms of $\gamma(t)$ corresponding to different predominant dephasing mechanism occur, however, typically $\gamma(t)$ can be written as $(t/T_2)^n$ for $n=1,2~\mathrm{or}~3$ \cite{Tisler.others2009FluorescenceandSpin,Taylor.others2008High-sensitivitydiamondmagnetometer}.

\begin{figure}[hbtp]
  \centering
  \includegraphics[width=0.9\linewidth]{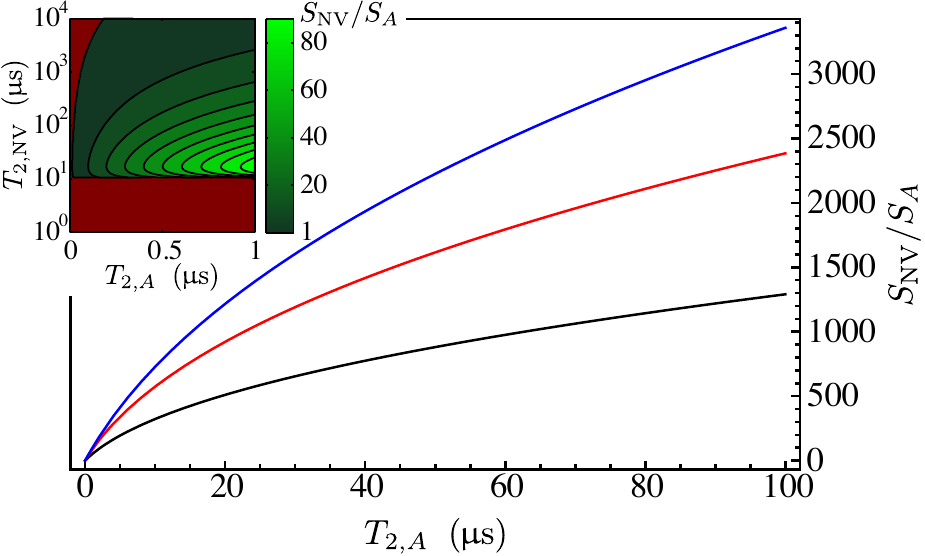}
  \caption{(Color online) Ratio $S_{\text{NV}}/S_A$ as a function of the amplifier spin coherence time $T_{2,\text{A}}$ for $T_{2,\text{NV}}=\unit{100}{\micro}{\second}$ (black), \unit{30}{\micro}{\second} (red), and \unit{15}{\micro}{\second} (blue). This ratio gives the factor by which the amplified sensor outperforms a conventional NV centre sensor. Here $\gamma(t)=t/T_2$, but we note that the curves for $\gamma(t)=(t/T_2)^{2,3}$ look almost identical. Other parameters: $r_A=\unit{1}{\nano\meter}, r_{\text{NV}}=\unit{11}{\nano\meter},\Delta=\unit{10}{\nano\meter}, t_{p,\text{NV}}=\unit{1}{\micro\second}, t_{p,\text{A}}=\unit{1}{\micro\second}+\tau_{\pi}$. Inset: excessive line-broadening or a significant mismatch in the coherence times prevent a sensitivity enhancement in the red region  (see text).}
\label{fig:ratio}
\end{figure}

For our comparison, we consider the levels $\ket{0}$ and $\ket{1}$ of an NV centre in the middle of a nanocrystal with radius \unit{10}{\nano\meter} and a control and measurement duration of $t_{p,\text{NV}}\approx \unit{1}{\micro\second}$ per cycle. In contrast, using the electron spin on the surface of the augmented sensor entails an additional control overhead of about $\tau_{\pi}$, i.e.~$t_{p,A}\approx t_{p,\text{NV}} +\tau_{\pi}$. Figure~\ref{fig:ratio} shows the ratio $S_{\text{NV}}/{S_A}$ as a function of the amplifier coherence time $T_{2,\text{A}}$ for several illustrative values of $T_{2,\text{NV}}$. The red regions of the inset show the parameter space for which our protocol is not necessarily advantageous, either due to excessive line-broadening preventing a highly selective $\pi$-pulse on the transition $\ket{0}\ket{\downarrow}\leftrightarrow\ket{1}\ket{\downarrow}$, or because $T_{2,\text{NV}}$ is so much larger than $T_{2,\text{A}}$ that it more than compensates the benefit of the amplifier spin. Figure~\ref{fig:ratio} impressively demonstrates that the magnetic moment sensitivity can be enhanced by up to three orders of magnitude for $T_{2,\text{NV}} \approx T_{2,\text{A}}$ with realistic parameters. Further improvements may be feasible by substituting the electron spin amplifier with a high spin system or molecular magnet. 

\begin{figure}[hbtp]
  \centering
  \includegraphics[width=0.8\linewidth]{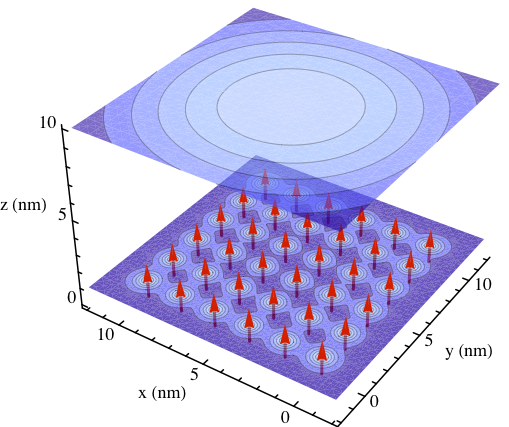}
  \caption{(Color online) The $z$-component of the magnetic field of a regular array of protons calculated in two planes $z=\unit{1}{\nano\meter}$ and $z=\unit{10}{\nano\meter}$. The $36$ dipoles with the magnetic moment of a proton are placed in the $z=0$ plane, pointing into the $z$-direction as indicated by the red arrows. The field range of the contour plots spans  $\unit{43.2} \nano \tesla$ in the  $z = \unit{10}{\nano\meter}$ plane and $\unit{2.61} \micro \tesla$ for $z=\unit{1}{\nano\meter}$. Individual dipoles are easily resolved at $z = \unit{1}{\nano\meter}$, but at a distance of $\unit{10}{\nano\meter}$ their field looks similar to that of a single stronger magnetic moment. Spatially resolving individual spins at this distance requires a sensitivity that goes far beyond the amount of detail reflected in this plot. }
  \label{fig:SpatialResolutionContour}
\end{figure}

We now turn to another important consideration: Given a sensor that can resolve magnetic fields at the level of the single Bohr or even nuclear magneton, its spatial resolution is vitally important for many applications such as mapping the constituent molecules and nuclei of complex organic molecules. Unsurprisingly, it is highly beneficial to be able to move very close to the sample in order to reliably distinguish individual magnetic moments. Figure~\ref{fig:SpatialResolutionContour} shows the $z$-component (i.e. the measured component) of the magnetic field for a regular array of  protons at two different heights above the surface. If the probe sample separation is of order the distance between individual dipoles, the nuclear spins  can be unambiguously resolved (given the required field sensitivity). On the other hand, the field is almost indistinguishable from that of a single stronger dipole if the sensor is scanning at $z=\unit{10}{\nano\meter}$, even with the ability to sense small magnetic fields of order $\unit{1}{\nano\tesla}$.

In summary, we have proposed a practical way of enhancing NV centre based magnetic field sensor with a pre-amplifier system. Such an augmented sensor possesses a magnetic moment sensitivity that is capable of resolving individual protons, both in terms of their field strength as well as spatially on an atomic scale. The sensor consists of two essential parts: the NV centre for optical readout coupled to an entirely spin-based system for the magnetic field detection in close proximity to the sample. Despite this partitioning of tasks, the fundamental complexity of this improved sensor design is similar to that of existing proposals and implementations.

\textit{Acknowledgements -} We thank Brendon Lovett and Stephanie Simmons for fruitful discussions. This work was supported by the National Research Foundation and Ministry of Education, Singapore, the DAAD, the Royal Society, St John's College and Linacre College, Oxford.


%

\end{document}